# Magnetic droplet solitons generated by pure spin currents

B. Divinskiy[1*], S. Urazhdin[2], V. E. Demidov[1], A. Kozhanov[3,4], A. P. Nosov[5], A. B. Rinkevich[5], and S. O. Demokritov[1,5]

[1]*Institute for Applied Physics and Center for Nanotechnology, University of Muenster, Corrensstrasse 2-4, 48149 Muenster, Germany*

[2]*Department of Physics, Emory University, Atlanta, GA 30322, USA*

[3]*Department of Physics and Astronomy, Georgia State University, Atlanta, GA 30303, USA*

[4]*National Research Nuclear University MEPhI, 115409 Moscow, Russia*

[5]*Institute of Metal Physics, Ural Division of RAS, Yekaterinburg 620041, Russia*

Magnetic droplets are dynamical solitons that can be generated by locally suppressing the dynamical damping in magnetic films with perpendicular anisotropy. To date, droplets have been observed only in nanocontact spin-torque oscillators operated by spin-polarized electrical currents. Here, we experimentally demonstrate that magnetic droplets can be nucleated and sustained by pure spin currents in nanoconstriction-based spin Hall devices. Micromagnetic simulations support our interpretation of the data, and indicate that in addition to the stationary droplets, propagating solitons can be also generated in the studied system, which can be utilized for the information transmission in spintronic applications.

Keywords: spin current, magnonics, magnetic nanostructures

# I. INTRODUCTION

Magnetic solitons are strongly nonlinear localized particle-like excitations in magnetic materials [1]. Among different types of solitons, dynamical solitons are particularly important. In addition to behaving like particles, they exhibit inherent high-frequency dynamics, making them attractive for applications associated with high-rate data transmission and processing of information. Dynamical magnetic envelope solitons, similar to the well-known temporal optical solitons [2], were observed about three decades ago in low-damping films of Yttrium Iron Garnet (YIG) [3,4], and have been intensively studied over the last years (see [5] and references therein). Because of the dynamic nature of these states, they are inhibited by the magnetic damping. In nanoscale magnetic systems, damping is generally larger than in the macroscopic YIG samples by more than an order of magnitude. As a result, the dynamical solitonic phenomena are strongly suppressed [6]. With the discovery of the spin transfer torque (STT) [7,8] it became possible to suppress the effective dynamical damping in magnetic nanosystems, enabling the observation of several types of dynamical magnetic solitons, including the dynamical skyrmions [9,10], the standing spin-wave bullets [11,12], and the magnetic droplets [13,14].

Magnetic droplet soliton is the dissipative counterpart of the magnon drop predicted for uniaxial ferromagnets with negligible damping, where easy-axis magnetic anisotropy leads to an attractive force between magnons [1,15]. Magnon condensation in the absence of damping results in the formation of a "drop" – a soliton-like object with the direction of magnetization in its core opposite to that in the surrounding volume. One of the remarkable features of this state is that all the spins at the boundary of the drop's core experience microwave-frequency precession around the anisotropy axis, preventing the drop from collapsing. Stable magnon drops cannot form in the equilibrium state of real magnetic materials, due to the non-zero damping. However, STT produced by the injection of spin current into a magnetic material allows one to locally compensate the effective damping, enabling the formation of stable magnetic droplets [13]. In contrast to the conservative magnon drops, magnetic droplets are strongly dissipative excitations stabilized by the balance between STT and natural damping. The possibility to generate magnetic droplets by STT was experimentally confirmed for the nanocontact spin torque nano-oscillators (NC-STNOs) based on the magnetic multilayers with perpendicular anisotropy (PMA) of the active layer [14].

To date, all the experimental studies of the magnetic droplet solitons have been performed in conventional NC-STNOs driven by spin-polarized electric current [14,16-21], where the damping is compensated only locally, and therefore the generated solitons are strongly localized in the region of the nanocontact. This approach does not allow one to explore propagating droplets, which can be particularly useful for the information transmission in applications utilizing droplets as information carriers. Instead of the magnetic multilayer nanocontacts, STT can be also produced in magnetic nanostructures by the spin-Hall effect (SHE) [22,23]. The geometry of SHE devices allows one to generate STT over spatially extended regions of magnetic materials, which is necessary for the stabilization of the propagating droplets. Over the last few years, a variety of SHE-based STT devices with novel geometries and functionalities have been demonstrated [24]. In particular, it was shown that magnetization dynamics can be excited by SHE [25,26], which led to the development of nano-oscillators driven by pure spin currents, which are characterized by moderate heat generation, high oscillation coherence, and efficient tunability of the oscillation frequency [27-33].

Two types of localized dynamic solitons, the dynamical skyrmions and the standing spin-wave bullets, have been recently observed in SHE oscillators based on planar point contacts [10,25]. However, because of the strongly localized injection of spin current in these devices, the solitons were found to be pinned in the active device area, similarly to the conventional NC-STNOs. Another type of SHE nanooscillators that received a significant attention due to their simplicity, geometric flexibility, and efficient operation, are based on a magnetic nanoconstriction [27,30-32]. These devices have been recently utilized to demonstrate

SHE-driven magnetization auto-oscillations in magnetic films with PMA [32]. However, only non-solitonic dynamics was observed.

Here, we report an experimental study of magnetization dynamics in nanoconstriction PMA spin Hall oscillators driven by strong currents. We show that a new dynamical state is formed at sufficiently large currents. This state is characterized by the oscillation at a frequency far below the ferromagnetic resonance. We use micromagnetic simulations to identify this regime with the formation of a magnetic droplet soliton. The spatial extent of the droplets generated in the nanoconstriction devices is larger than in the previously studied systems, and they exhibit complex spatio-temporal dynamics. The simulations quantitatively agree with the experimentally observed properties of droplets, and predict the possibility to generate propagating magnetic droplets stabilized by the spatially extended injection of spin current.

## II. EXPERIMENT

The studied devices are based on the Pt(5)/[Co(0.2)/Ni(0.8)]$_4$/Ta(3) magnetic multilayer with PMA. Here, thicknesses are given in nm. The multilayer is patterned into a bowtie-shaped nanoconstriction with the width of 100 nm, the opening angle of 22°, and the radius of curvature of about 50 nm [Fig. 1(a)]. The operation of the device relies on the SHE in Pt, which converts the dc electrical current $I$ flowing in the plane of the multilayer into spin current $I_s$ flowing out of plane, with spin polarization $\sigma$ oriented in-plane, perpendicular to the direction of current $I$. The spin current is injected into the Co/Ni multilayer, exerting STT on its magnetization. The effects of STT result in either a decrease or an increase of the effective magnetic damping, depending on the direction of spin current polarization $\sigma$ relative to the magnetization **M** [34]. In the former case, the magnetic damping can be completely compensated by a sufficiently large spin current, resulting in the dynamical instability that leads to magnetic auto-oscillations.

In the studied device geometry, the abrupt narrowing of the Pt layer in the nanoconstriction region causes a strong local increase of the electric current density $J$ [Fig. 1(b)]. Since the spin current injected into the Co/Ni multilayer is proportional to the current density in Pt, this region of large current density defines the active device area, where STT is sufficiently strong to completely compensate the damping and cause the magnetization instability. While the current density rapidly decreases with increasing distance from the nanoconstriction, considerable STT effects are expected outside the active device area. For instance, the current density at the distance of 1 µm from the nanoconstriction center is about 5-7% of that at the center. Thus, at currents significantly above the instability threshold, damping can be almost completely compensated for a relatively large region of the magnetic film.

According to the symmetry of SHE, the polarization **σ** of spin current injected into Co/Ni is parallel to the film plane. If the magnetization **M** is oriented normal to the film plane, the effect of STT on damping vanishes. Therefore, to achieve current-induced instability in the studied Co/Ni film with PMA, we apply an in-plane static magnetic field $H_\parallel$=1000-2000 Oe to tilt the magnetization with respect to the film normal. Additionally, to prevent magnetization switching by SHE, we apply a small constant out-of-plane magnetic field $H_\perp = 200$ Oe. Under these conditions, the angle $\theta$ between **σ** and **M** [see Fig. 1(a)] is estimated to vary between 57° and 74°, depending on the magnitude of $H_\parallel$.

To study the effect of spin current on the magnetization of the Co/Ni multilayer, we use the micro-focus Brillouin light scattering (BLS) spectroscopy technique [35]. We focus the probing laser light with the wavelength of 532 nm and the power of 0.25 mW into a diffraction-limited spot on the surface of the Co/Ni multilayer [Fig. 1(a)], and analyze the spectrum of light inelastically scattered from the magnetization oscillations. The detected signal – the BLS

intensity – is proportional to the intensity of the oscillations at the selected frequency, permitting spectral analysis of the magnetization oscillations in the sample.

Figures 2(a), (b) show the BLS spectra measured at $H_\parallel = 2000$ Oe, with the probing laser spot positioned at the center of the nanoconstriction. At $I>2$ mA, a narrow intense peak appears in the spectra [Fig. 2(a)], marking an onset of single-mode current-induced magnetization auto-oscillations caused by the complete compensation of damping by the spin current. The oscillation frequency is close to the frequency of the uniform ferromagnetic resonance (FMR) in the Co/Ni multilayer [32]. The auto-oscillation peak rapidly grows with increasing current, reaching a maximum intensity at $I = 3.1$ mA. At larger currents, the peak starts to broaden, while its intensity decreases. Simultaneously, broad noise-like spectral features emerge at frequencies below 3 GHz [see Fig. 2(a) for $I=3.5$ mA], indicating that the system transitions to a new regime characterized by complex magnetization dynamics. At currents above 3.5 mA (Fig. 2(b)), this broad spectrum evolves into a well-defined low-frequency (LF) oscillation peak with the center frequency of about 2 GHz, while the peak corresponding to the high-frequency (HF) mode gradually disappears. Note that the frequency of the LF peak is almost current-independent or slightly increases with increasing current, whereas the central frequency of the HF peak decreases with current.

Figure 2(c) shows the current dependencies of the peak intensity for the HF (open symbols) and the LF (solid symbols) modes. As seen from these data, the decrease in the intensity of the high-frequency oscillations is clearly correlated with the increasing intensity of the low-frequency mode, indicating that the two modes compete for the angular momentum delivered into the magnetic system by the spin current. Meanwhile, the total BLS intensity integrated over the spectrum monotonically increases with increasing $I$ over the entire studied range of current (Fig. 2(d)). In contrast, in films with in-plane magnetization, the total intensity of current-induced oscillations is strongly suppressed at large currents by the nonlinear magnon-magnon scattering processes, which are facilitated by the degeneracy of the magnon spectrum [24]. This degeneracy is lifted in the Co/Ni multilayers with PMA, enabling very large amplitudes of the current-induced magnetization precession [32], and allowing a complete local reversal of the magnetization necessary for the formation of droplet solitons.

### III. MICROMAGNETIC SIMULATIONS

To elucidate the nature of the observed low-frequency dynamical mode, we perform micromagnetic simulations using the MuMax3 software [36]. The simulation model comprises the area of the magnetic layer centered on the nanoconstriction, discretized into $4 \times 4 \times 4$ nm$^3$ cells. The effect of spin current is modeled by the Slonczewski's STT term [7]. The magnitude of STT is assumed to be proportional to the current density in Pt. Its distribution is calculated for the studied structure by using the COMSOL Multiphysics software [Fig. 1(b)]. We also take into account the non-uniform Oersted field of the current, determined from the calculated current distribution. The magnetic parameters of the Co/Ni multilayer used in the simulations – the saturation magnetization $M_s = 490$ kA/m and the anisotropy coefficient $K_u = 0.23$ MJ/m$^3$ – were determined from separate measurements utilizing vibrating-sample and magneto-optical Kerr-effect magnetometries. We assume the standard values $\alpha = 0.02$ for the Gilbert damping constant of Co/Ni, and $\theta_{SH} = 0.08$ for the effective spin-Hall angle of Pt. All the simulations are performed at zero temperature, with the same out-of-plane magnetic field $H_\perp = 200$ Oe as in our experiments.

Figure 3(a) illustrates the calculated time dependence of the normalized out-of-plane component of magnetization $m_z = M_z / M_s$ at $H_\parallel = 2000$ Oe, with current $I=5$ mA applied starting at time $t=0$. Under these conditions, STT-induced dynamical instability develops on the sub-nanosecond time scale. The component $m_z$ initially starts to oscillate with the frequency close to the FMR frequency and with a rapidly growing amplitude, and completely reverses at $t$ slightly above 0.6 ns. This instability is observed at $I>I_c\approx 4.5$ mA, in a reasonable agreement with the characteristic currents in the experiment. As the current is reduced towards $I_c$, it takes an increasingly long time for the instability to develop, resulting in quasi-stable current-induced oscillations at frequencies close to the FMR, consistent with the behaviors observed in our measurements at small currents [see Fig. 2(a)]. The weak instability at currents close to $I_c$ likely becomes completely suppressed in our experiment by thermal fluctuations neglected in the simulations. Below, we focus on the large-current behaviors that are not expected to be significantly affected by thermal effects.

Figure 3(b) shows a snapshot of the spatial distribution of magnetization after its complete reversal, at $I=5$ mA and $t=0.625$ ns. Arrows in this map represent the projection of the magnetization on the film plane, while the color represents its out-of-plane component $m_z$. The magnetization distribution is typical for a droplet soliton. The magnetization is nearly completely reversed in the droplet core, as can be also seen from Fig. 3(c) showing the profile of $m_z$ across the section indicated in Fig. 3(b) by a dashed line. At the boundary of the core, all the magnetic moments are aligned in the same direction. The droplet is shifted to one of the edges of the nanoconstriction, due to the effect of the Oersted field of the driving current. A similar edge droplet was predicted for narrow magnetic nanowires [37].

Snapshots obtained at different instants of time (Fig. 4) illustrate the spatio-temporal dynamics of the nucleated droplet. As seen from Figs. 4(a)-4(c), the droplet initially increases in size, while the magnetization at its boundary precesses in the counterclockwise direction. The expansion of the droplet is caused by the effective spin-Hall field [38] produced by STT. This field is known to result in either expansion or shrinking of magnetic domains, depending on the relative orientation of the magnetization in the domain walls and the polarization $\sigma$ of the spin current [39]. Therefore, the observed initial expansion of the droplet is produced by the same mechanism as the expansion of topologically trivial magnetic bubble domains driven by STT [40]. The droplet is distinguished from the bubble domains by its dynamical nature. Because of the precession of the magnetic moments at the droplet boundary, the effective spin Hall field is time-dependent. As a result, the initial expansion of the droplet ends at $t\approx0.850$ ns, when the magnetic moments at its boundary align approximately parallel to $\sigma$, and the effective spin-Hall field vanishes [Fig. 4(c)]. Further rotation of the magnetic moments leads to the inversion of the sign of the effective field, and the droplet starts to shrink [Figs. 4(c)-4(e)]. If the magnetic moments in the droplet boundary precessed perfectly in-phase, the droplet would experience periodic "breathing" with a well-defined frequency, as observed in the nanocontact droplet oscillators [14]. In the studied nano-constriction devices, the droplet dynamics is significantly more complex, as illustrated in Figs. 4(e)-4(h). The dynamical complexity is associated with the inhomogeneous distribution of the effective field, which leads to the spatial variation of the precession phase, as can be clearly seen from the increasingly significant variations among the directions of arrows in Figs. 4(c)-4(h). Because of these variations, different parts of the droplet experience different forces from by the spin-Hall field, which perturbs the shape of the droplet, and results in complex spatio-temporal dynamics. This observation is in agreement with the experimental data, which show that at large driving currents, the spectra of the current-induced magnetization oscillations become very broad.

To correlate the results of the simulation with the experiment, we calculate the Fourier spectrum of the time dependence of the spatially averaged out-of-plane magnetization component $m_z$ (Fig. 5). Despite the dynamical complexity, the calculated spectrum exhibits a pronounced peak, which is remarkably similar to the BLS spectrum measured at $I = 4$ mA [see Fig. 2(b)]. In particular, its central frequency 2.30 GHz is very close to the experimental value 2.25 GHz. Additional simulations for different currents show that the frequency is nearly independent of current, in agreement with the data. A good agreement of our simulations with the experimental observations allows us to unambiguously conclude that the observed low-frequency mode is associated with the formation of a droplet soliton.

Analysis of the dependence of the low-frequency spectrum on the magnitude of the static in-plane magnetic field $H_{\parallel}$ provides further support for this conclusion. The observed spectra of the low-frequency oscillations strongly depend on $H_{\parallel}$, as illustrated in Fig. 6(a) for $I$=4 mA. In particular, the measured central peak frequency linearly increases from 1.3 GHz at $H_{\parallel}$=1000 Oe to 2.25 GHz at $H_{\parallel}$=2000 Oe [open symbols in Fig. 6(b)]. The observed linear dependence is consistent with the behaviors expected for the droplet soliton [14]. The dependence of the central peak frequency on field obtained from the simulations, solid symbols in Fig. 6(b), is in a good agreement with the experiment over the entire studied range of $H_{\parallel}$, supporting our interpretation of the observed behaviors.

We now discuss the effects of the spatially extended injection of spin current produced by the SHE. Since the droplet solitons are stabilized by STT, their properties are strongly dependent on the spatial distribution of the spin current injection. In the nanocontact spin-torque devices, STT is exerted only over the area of the nanocontact. As a result, the droplets generated in the nanocontacts exhibit a nearly circular shape, with the dimensions close to the diameter of the nanocontact [18,21]. In contrast, in the nanoconstriction SHE devices, the spin current density does not abruptly vanish outside of the nanoconstriction area. The resulting spatial dimensions of the droplets can be as large as 200-300 nm in the nanoconstriction with the width of 100 nm used in our work. As shown above, this large size results in complex dynamics and an irregular shape of the droplet, which originate from the non-uniform distribution of the magnetization precession phase at its boundary.

We emphasize that the spatial distribution of the precession phase can significantly affect the spatial characteristics of the droplet. For the uniform phase distribution, the effective spin-Hall field produces periodic expansion and shrinking of the droplet, as discussed above.

However, if the precession phases are opposite at the opposing points at the droplet boundary, the spin-Hall field is expected to cause translational motion of the droplet, as was also shown for the magnetic skyrmions [40] where the direction of the magnetization rotates around the soliton boundary due to the Dzyaloshinskii-Moriya interaction (DMI) [41,42].

In the studied system, the effects of DMI are negligible, so the stationary phase distribution characteristic of skyrmions is generally not expected. Nevertheless, our simulations indicate that non-trivial phase distributions enabling the SHE-induced translational motion of droplets can spontaneously form at sufficiently long times of the order of tens of nanoseconds. Figure 7(a) shows the spatial magnetization map obtained at $t$=36.050 ns, with $I$=6 mA applied starting at $t$=0. In addition to the irregular-shaped droplet in the nanoconstriction region, a small circularly shaped droplet with dimensions of about 50 nm is located outside of the active area. The magnetization is pointing to the left at the left boundary of the small droplet, and to the right at its right boundary [inset in Fig. 7(a)]. Consequently, the droplet is expected to experience a net effective spin Hall field driving its directional motion, as confirmed by the snapshots in Figs. 7(b)-7(d) for later instants of time. As seen from these data, the droplet is indeed moving to the left while shrinking in size, and is finally annihilated at the distance of more than 1.5 μm from the nanoconstriction center. We emphasize that the moving droplet remains stable due to the non-vanishing STT over its entire propagation path in the studied nanoconstriction device. While at large distances from the active area the magnitude of STT becomes too small to completely compensate the natural damping, it is still sufficiently large to maintain the stability of the nucleated droplet over a significant propagation distance.

The generation of propagating droplets can be important for the implementation of novel devices for transmission and processing of information on the nanoscale. The controllability of the droplet generation process can be improved by utilizing nanopatterned material systems where a combination of DMI and geometric boundary effects favors the spatial variation of the precession phase necessary for the extraction of the droplet from the nucleation area, and for its directional motion [43]. By optimizing the geometry of the samples, one can also control the magnitude and the distribution of STT outside of the active nanoconstriction area, to increase the propagation distance of the generated droplets.

## IV. CONCLUSIONS

In conclusion, we have experimentally demonstrated that magnetic droplet solitons can be efficiently nucleated and sustained by spin currents produced by the spin Hall effect. Micromagnetic simulations reveal complex dynamics of the magnetic droplets under the influence of SHE, and predict that propagating droplet solitons can be generated in the studied devices. Our findings indicate that, in addition to the generation of droplet solitons, pure spin currents can also be used to control their propagation in magnetic films. Our observations open new possibilities for the application of dynamical magnetic solitons in the future nanoelectronic devices.

## ACKNOWLEDGMENTS

This work was supported in part by the Deutsche Forschungsgemeinschaft, the NSF Grant Nos. ECCS-1509794 and DMR-1504449, and the program Megagrant № 14.Z50.31.0025 of the Russian Ministry of Education and Science.

**Figure captions**

FIG. 1. (a) Schematic of the experiment. (b) Calculated distribution of the driving current density in the plane of the device.

FIG. 2. (a) and (b) BLS spectra of current-induced magnetization oscillations, measured at the labeled values of the driving current. (c) Peak intensities of the high-frequency (HF) and the low-frequency (LF) modes vs current $I$. (d) Current dependence of the intensity integrated over the BLS spectrum. The data were obtained at $H_\parallel = 2000$ Oe.

FIG. 3. (a) Calculated time dependence of the normalized out-of-plane magnetization component $m_z$, demonstrating the development of the current-induced instability. The current is applied starting at $t=0$. (b) Calculated spatial magnetization map at $t=0.625$ ns, shortly after the droplet is nucleated. Arrows represent the in-plane magnetization component, and the colors represent its out-of-plane component $m_z$. (c) Profile of $m_z$ along the dashed line in panel (b). Simulations were performed at $I = 5$ mA and $H_\parallel = 2000$ Oe.

FIG. 4. Temporal evolution of the nucleated droplet soliton, at the labeled instants of time $t$. The driving current is applied starting at $t=0$. The arrows represent the in-plane magnetization component and the colors represent its out-of-plane component. The simulations were performed at $I = 5$ mA and $H_\parallel = 2000$ Oe.

FIG. 5. Fourier spectrum of the time dependence of the spatially averaged out-of-plane magnetization component $m_z$, for the dynamics shown in Fig. 4. Vertical dashed line indicates the central frequency of the spectral peak.

FIG. 6. (a) BLS spectra of magnetic oscillations in the droplet mode, measured at $I = 4$ mA, at the labeled magnitudes of the magnetic field $H_\parallel$. (b) Dependence of the central frequency of the spectral peak on $H_\parallel$, determined from the measurements (open symbols) and from the micromagnetic simulations (solid symbols). Lines are linear fits of the data.

FIG. 7. Calculated magnetization maps demonstrating the generation of a propagating droplet soliton, at the labeled instants of time *t*. The driving current is applied starting at *t*=0. The colors represent the out-of-plane magnetization component. Inset in (a) shows a magnified view of the propagating droplet. The simulations were performed at $I = 6$ mA and $H_{\parallel} = 1000$ Oe.

(a)

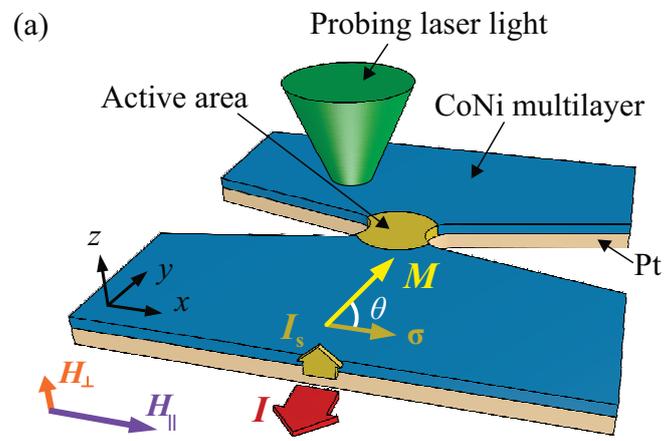

(b)

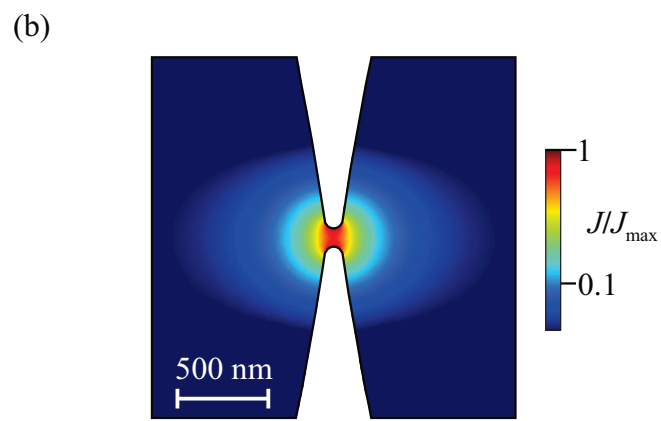

Fig. 1

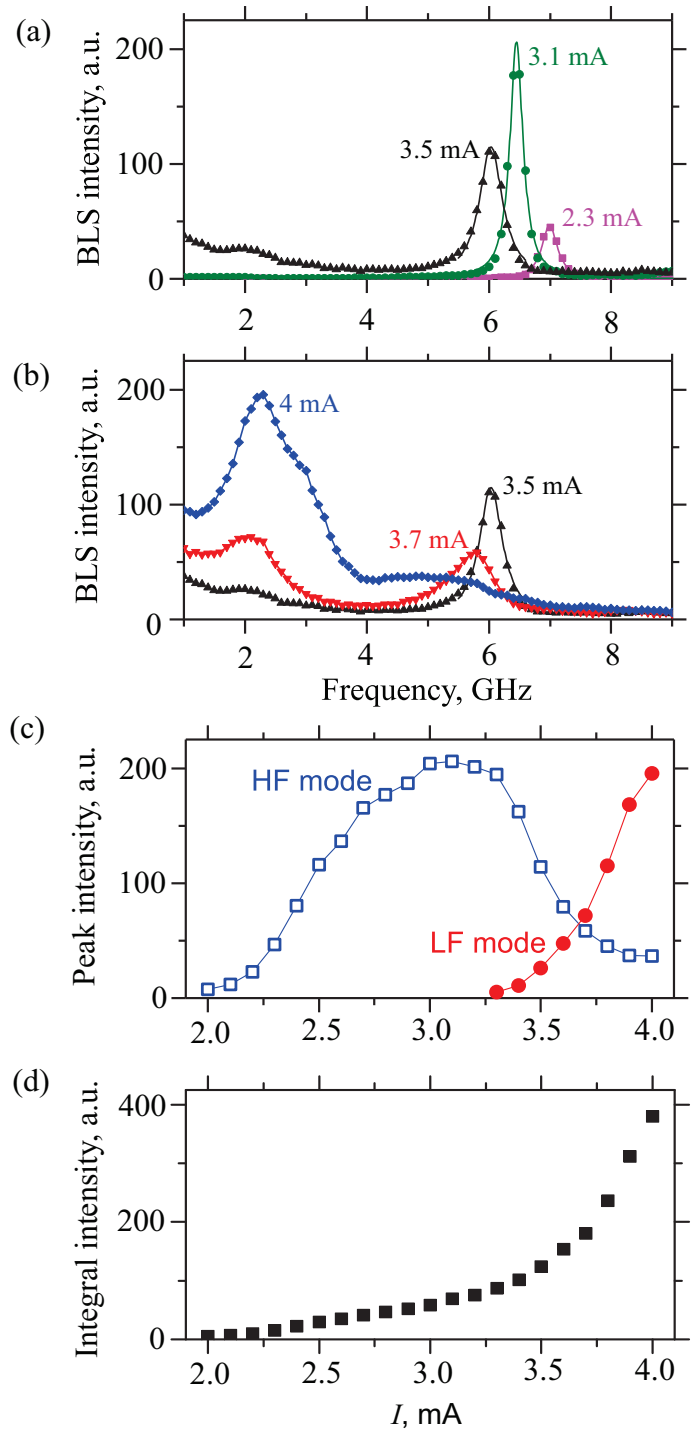

Fig. 2

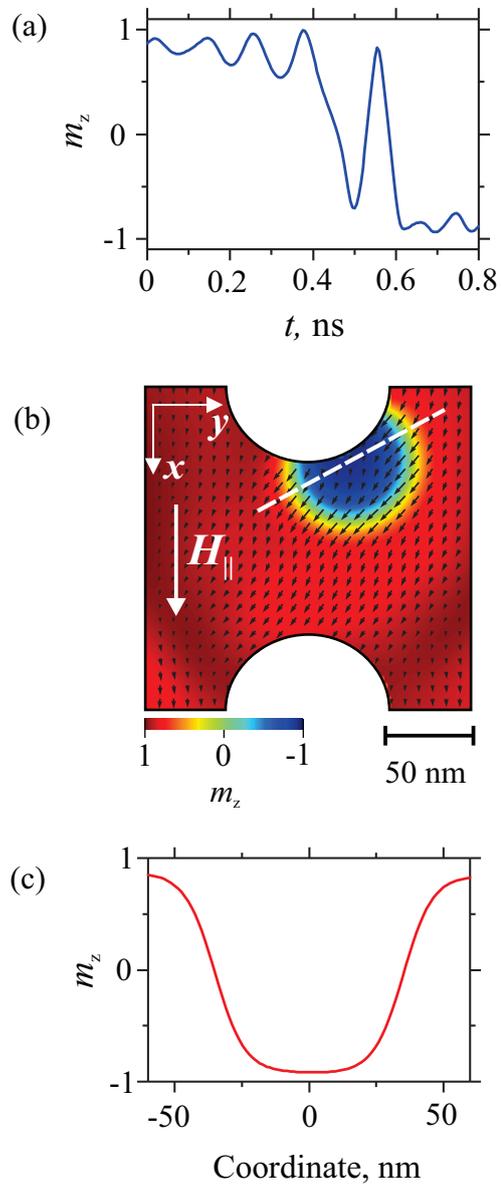

Fig. 3

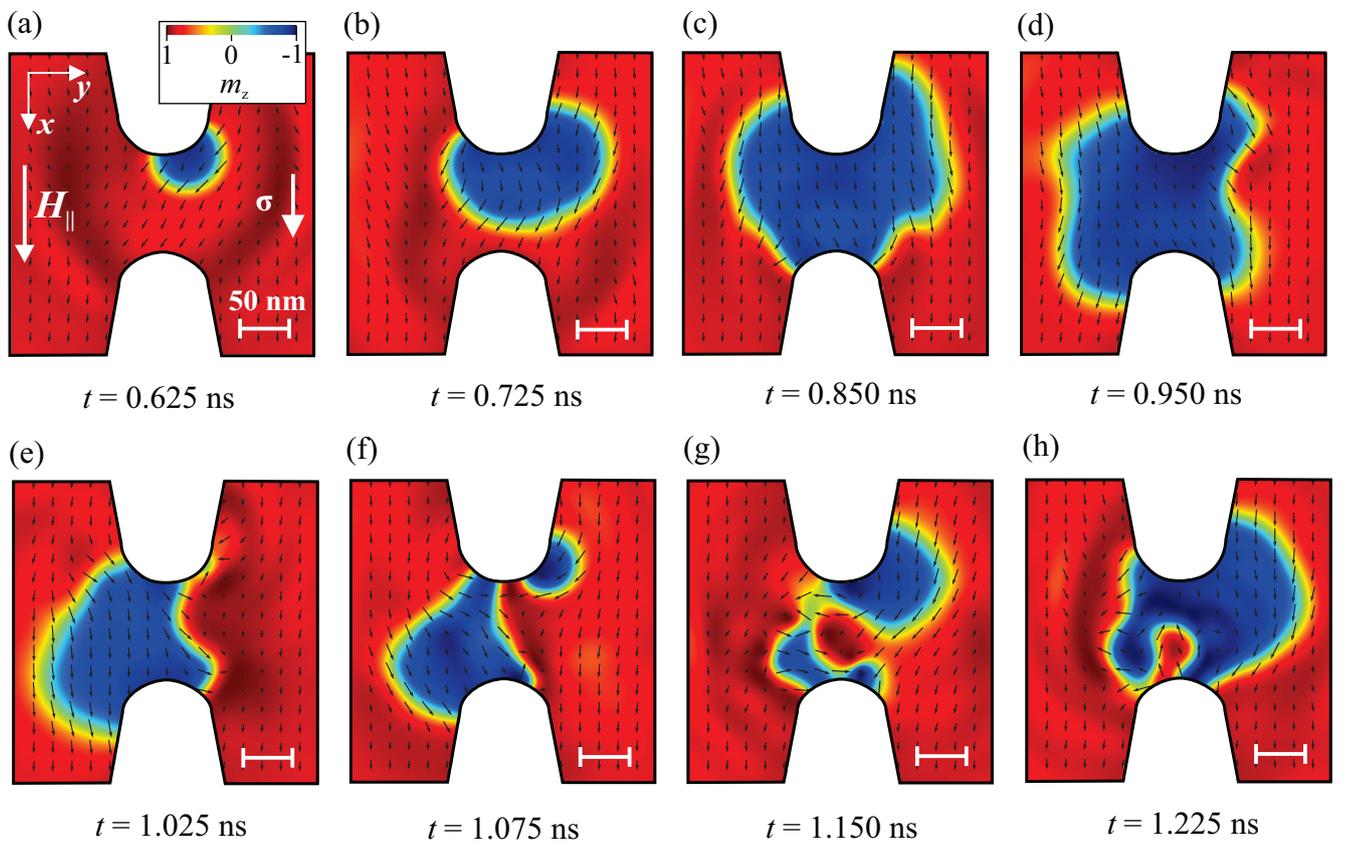

Fig. 4

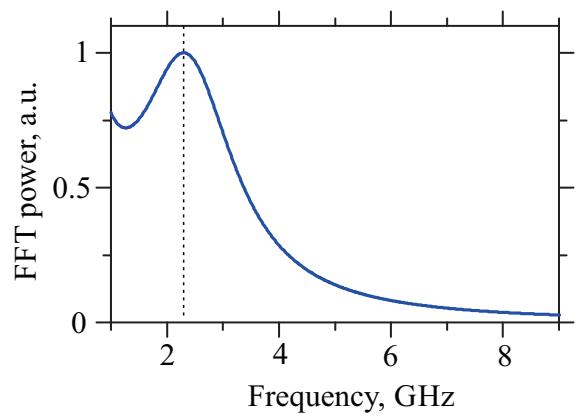

Fig. 5

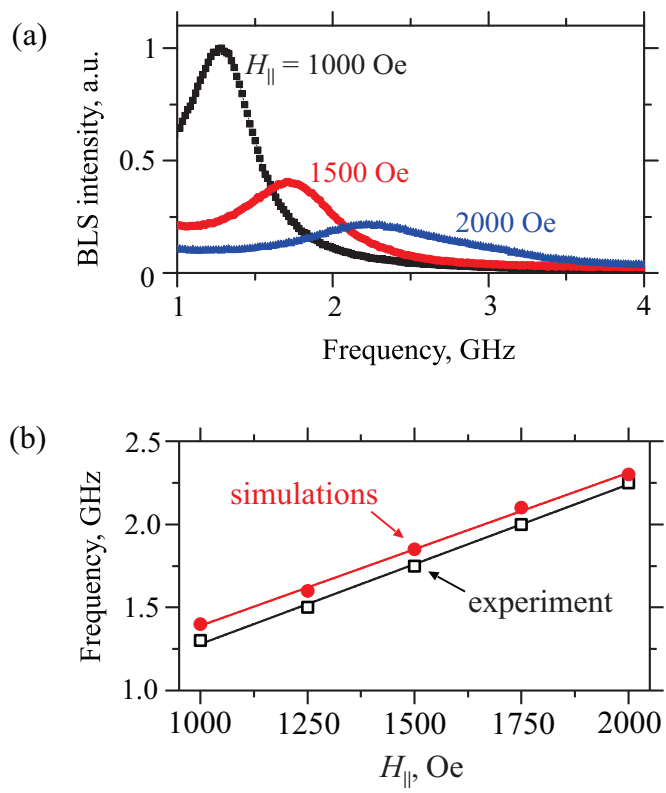

Fig. 6

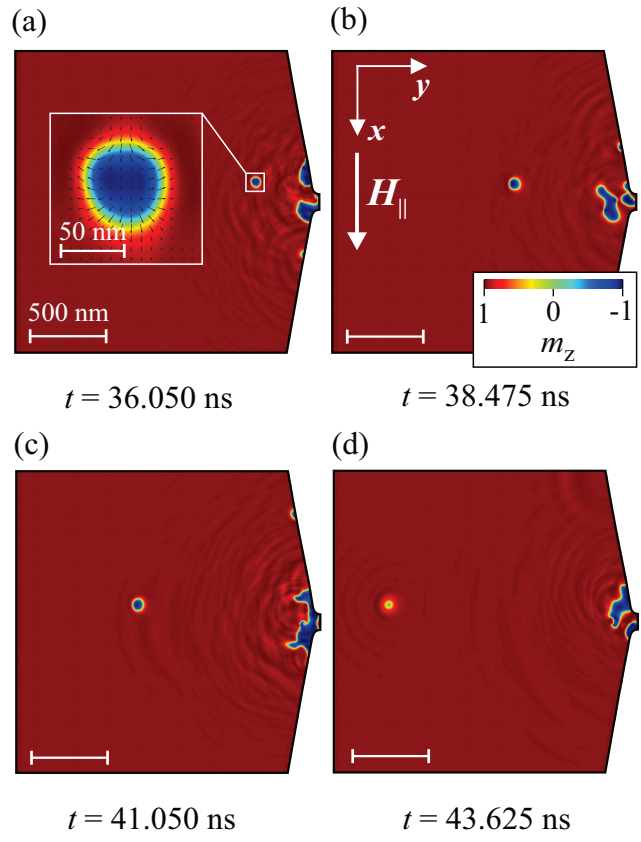

Fig. 7